\documentclass[twocolumn,trackchanges]{aastex701}
\usepackage{amsmath}
\hypersetup{linkcolor=red,citecolor=blue,filecolor=cyan,urlcolor=magenta}

\graphicspath{{./}{figures/}}

\begin{document}

\title{Free-floating Planets Produced by Planet-Planet Scatterings: Ejection Velocity and Survival Rate of Their Moons}

\author{Xiumin Huang}
\affiliation{Tsung-Dao Lee Institute, Shanghai Jiao-Tong University, Shanghai, 201210, China}
\email[show]{xm\_huang@sjtu.edu.cn}  

\author{Dong Lai}
\affiliation{Tsung-Dao Lee Institute, Shanghai Jiao-Tong University, Shanghai, 201210, China}
\affiliation{Center for Astrophysics and Planetary Science, Department of Astronomy, Cornell University, Ithaca, NY 14853, USA}
\email[show]{donglai@sjtu.edu.cn}


\begin{abstract}
The discovery of numerous free-floating planets (FFPs) has intensified interest in their origins and dynamical histories. A leading formation mechanism is planet–planet scatterings in unstable multi-planetary systems, which can naturally lead to planetary ejections. If these planets originally host moons, it remains an open question whether such satellites can remain gravitationally bound to FFPs after ejection. In this work, we investigate both the ejection velocity of FFPs produced by planet–planet scatterings and the survival rate of their potential moons; we estimate the latter by determining the statistics of the minimum planet-planet distance prior to planet ejection, and comparing it to the initial orbital radius of the moon relative to its host planet.  Using the circular restricted three-body framework, we derive an analytical boundary for the ejection velocity based on Jacobi energy conservation, which agrees with the results of integrations. We also identify a minimum planetary mass required for successful ejection.  For two-planet systems with finite planetary masses, we use simulations and analytical arguments to determine how the ejection velocity scales with the planetary mass and initial semi-major axis. We contrast our results to the ejection of planets around binaries in unstable orbits. Extending our analysis to three-planet systems yields similar results, reinforcing the robustness of our conclusions. These findings offer insights into the property of FFPs and inform future efforts to search for exomoons around them. 
\end{abstract}

\section{Introduction}
Free-floating planets (FFPs) are planetary-mass objects that are not gravitationally bound to any host star. Over the past decades, numerous FFPs have been discovered in young star clusters through infrared surveys, direct imaging, and gravitational microlensing \citep{zapatero2000discovery,scholz2012substellar,ramirez2012new,liu2013extremely,mroz2017no,ryu2021kmt,miret2022rich,sumi2023free}. Their inferred masses span a wide range, from sub-Earth mass to tens of Jupiter masses \citep{sumi2023free}.

Several formation channels have been proposed to explain the origin of FFPs. One scenario involves the direct gravitational collapse of molecular cloud fragments, akin to the process of star formation \citep{padoan2002stellar,hennebelle2008analytical,luhman2012formation,miret2022rich}. This may produce FFPs with masses of a few Jupiter masses or higher. Another channel involves dynamical ejection triggered by interactions with massive bodies, such as stellar flybys or binary stars. For instance, close stellar encounters which are frequent in young and dense clusters, can disrupt planetary systems and lead to various outcomes, including FFPs and planet capture \citep{rodet2021correlation,rodet2022impact,yu2024free}. In circumbinary systems, interactions with the central binary can similarly result in planetary ejection \citep{sutherland2016fate,coleman2024properties}.

Among the various dynamical processes proposed, planet–planet scattering is recognized as a particularly efficient and natural mechanism for producing FFPs \citep{rasio1996dynamical,weidenschilling1996gravitational,ma2016free,li2021giant,bhaskar2025properties,hadden2025free}. In compact and dynamically unstable planetary systems, close encounters between planets can trigger a range of outcomes, including the ejection of one or more planets into the interstellar space, direct collisions between planets, or the inward scattering of a planet into the host star \citep{chambers1996stability,chatterjee2008dynamical,deck2013first,li2021giant}. Among these, ejection events naturally lead to the formation of FFPs. This mechanism is strongly supported by both numerical simulations and observational evidences. For instance, the observed eccentricity distribution of giant exoplanets can be reasonably reproduced by dynamical evolution driven by scatterings \citep{lin1997origin,ford2008origins,chatterjee2008dynamical,raymond2009planet,anderson2020situ,li2021giant}. Furthermore, planet–planet scatterings could act as a trigger for secondary dynamical processes, such as the von Zeipel–Lidov–Kozai mechanism, which further reshapes planetary system architectures over time \citep{nagasawa2008formation,beauge2012multiple,lu2025planet}.

The masses and abundance of FFPs are obviously important questions that relate to the origin of FFPs \citep{hadden2025free}. The velocities of FFPs also encode vital information about their formation history. FFPs originating from circumbinary systems and ejected by binary stars typically acquire higher ejection velocities than those produced through planet–planet scattering. \citet{coleman2024properties} demonstrated that such binary-ejected FFPs often move significantly faster than the local stellar background (see Section \ref{scbp} for our own simulation and scaling results). On the other hand, FFPs produced by planet-planet scatterings have ejection velocities that depend on their initial distances from the host star. In contrast, FFPs formed via gravitational collapse of molecular cloud fragments are expected to share similar velocities with nearby stars, as they inherit the kinematic properties of their natal environment. Therefore, characterizing the ejection velocity of FFPs offers a powerful diagnostic tool for constraining their origin.

When a planet is ejected from its parent system, a key question is whether any moons it originally hosted can remain gravitationally bound after the ejection. Such moons are considered potentially habitable environments, as tidal heating may sustain subsurface liquid water \citep{reynolds1987europa,scharf2006potential}, and the absence of a nearby host star reduces exposure to harmful stellar radiation. Numerical simulations of specific systems have explored the fate of moons during planet–planet scattering events \citep{debes2007survival,hong2018innocent,rabago2019survivability}, indicating that in some cases, moons can survive and remain bound to their host planets after ejection. Upcoming missions, including the Chinese Space Station Telescope and the Roman Space Telescope, are expected to improve our understanding of these systems and may enable the detection of both FFPs and their accompanying exomoons \citep{fu2025detecting,derocco2025reconstructing}.

In this paper, we investigate both the ejection velocity of FFPs produced by planet–planet scatterings and the likelihood that their moons remain gravitationally bound after ejection. We begin with the circular restricted three-body framework, treating one planet as a test particle, and derive an analytical expression for the ejection velocity based on the Jacobi energy conservation. This result shows excellent agreement with numerical simulations and reveals a minimum planetary mass required for successful ejections. We then extend the analysis to general two-planet systems with finite planetary masses, using $N$-body simulations and analytical arguments to examine how the ejection velocity depends on the planet mass and the initial semi-major axis. We contrast these result to the ejection of circumbinary planets. To assess the survivability of moons, we use the minimum separation between planets during close encounters as a proxy, assuming a moon remains bound if its orbital radius around the host planet is a few times less than the minimum planet-planet distance. Furthermore, we apply our framework to three-planet systems and demonstrate that the core results remain robust in more complex dynamical configurations.

This paper is structured as follows. Section \ref{s2} introduces the setup for producing FFPs in two-planet systems. In Section \ref{stp}, we analyze the ejection velocity within the circular restricted three-body framework, combining analytical derivations with numerical validation. Section \ref{sgc} extends the analysis to two-planet systems with finite masses, addressing both the inferred survival probability of moons and the scaling of ejection velocity. Section \ref{s3p} investigates the dynamics of three-planet systems and tests the robustness of our conclusions. Section \ref{scbp} examines the ejection velocities of circumbinary planets for comparison. Section \ref{scon} summarizes our findings and provides concluding remarks.

\section{Scenario of two-planet scatterings}\label{s2}
We consider two planets with masses $m_1$ and $m_2$ orbiting a star with mass $M$ in nearly circular and co-planar configurations. The stability criterion for such a two-planet system is \citep{gladman1993dynamics}
\begin{equation}
    {a_2} - {a_1} > 2\sqrt3 { R_{\rm H}},
\end{equation}
where $a_1$ and $a_2$ are semi-major axes of the inner and outer planets respectively, and $R_{\rm H}$ is the mutual Hill radius, given by
\begin{equation}
    {R_{\rm H}}{\rm{ = }}\frac{{{a_1} + {a_2}}}{2}{\left( {\frac{{{m_{12}} }}{{3M}}} \right)^{1/3}}  
\end{equation}
with $m_{12}=m_1+m_2$. So when initially 
\begin{equation}
    k \equiv \frac{{{a_2} - {a_1}}}{{{R_{\rm H}}}} < 2\sqrt 3,    
\end{equation}
a system is unstable. Such instability can lead to  planet–planet collisions, inward scattering of a planet into the host star, or ejection of a planet from the system. In the case of ejection, the unbound planet becomes a FFP. If the ejected planet originally hosted a moon, the fate of that moon depends on the strength of gravitational interactions during close encounters. A sufficiently close approach between planets may disrupt the planet–moon system and unbind the moon. However, if the encounters are not too close, the moon may remain bound and be ejected together with its host. In this work, we aim to characterize the velocity properties of the resulting FFPs, and to assess the likelihood that FFPs retain their moons after ejection.

To this end, we run the $N$-body code \texttt{REBOUND} \citep{rein2012rebound} with the \texttt{IAS15} integrator \citep{rein2015ias15}. Unless otherwise noted, the star is set to $M=1M_{\odot}$ in the simulations, the semi-major axis of the inner planet is $a_1=1$ au, and the inner planet is set to be more massive than the outer one. All bodies are treated as point masses, making the simulations scale-free. Under this assumption, physical collisions—either between planets or with the star—are excluded from the dynamical outcomes. Each system is initialized in an unstable configuration, ensuring that close encounters lead to ejection of the less massive planet, which becomes a FFP. While moons are not explicitly modeled, we estimate the survivability of a hypothetical moon by tracking the minimum separation $r_{12,\rm{min}}$ between the two planets prior to planet ejection. A moon around $m_2$ (with initial orbital radius $r_{\rm 2m}$) can survive the planet-planet encounters if its orbital timescale, $\sqrt{r_{\rm 2m}^3/(Gm_2)}$, is less than the timescale of the closest encounter, $\sqrt{r_{\rm 12,min}^3/(Gm_{12})}$, i.e.,
\begin{equation}\label{eq_r2m}
    r_{\rm 2m} < fr_{12,\min}\left(\frac{m_2}{m_{12}} \right)^{{1}/{3}},
\end{equation}
where $G$ is the gravitational constant and $f$ is of order unity. Similar condition applies to the moon around $m_1$. A high survival probability of the moon requires $f \sim 1/2$ (see \citealp{rodet2022impact} for the survival of a planet during a stellar flyby).

We adopt the following criterion to identify ejection events: the separation between the two planets exceeds 50 au ($r_{12} > 50$ au), and the orbital energy of the outer planet is positive ($E_2 > 0$). The energy of the outer planet per unit mass is given by
\begin{equation}
    {E_2} = \frac{1}{2}v_2^2 - \frac{{GM}}{{{r_{02}}}} - \frac{{G{m_1}}}{{{r_{12}}}},
\end{equation}
where $v_2$ is the velocity of the outer planet relative to the system’s center of mass, and $r_{02}$ is its distance to the central star. By energy conservation, the velocity of FFP at infinity is 
\begin{equation}\label{eq_vinf}
    {v_{\infty }} = \sqrt {2{E_{\rm f}}},
\end{equation}
where $E_{\rm f}$ is the final energy of the outer planet (per unit mass) after ejection. During each simulation, we record the minimum planet–planet separation prior to ejection to assess the survivability of a hypothetical moon, and compute $v_\infty$ to quantify the ejection velocity of the resulting FFP.

\section{Two-planet scatterings: test particle case}\label{stp}
\subsection{Analysis}
When the outer planet is a test particle (i.e., $m_2 \to 0$), the system reduces to that of a circular restricted three-body problem (CR3BP). We adopt normalized units such that $G(M + m_1) = 1$ and $a_1 = 1$, and let $\mu = m_1 / (M + m_1)$. In the co-rotating frame, where the $x$-axis always points from the central star $M$ to the inner planet $m_1$, the coordinates of the star and the inner planet are fixed at $(-\mu, 0, 0)$ and $(1 - \mu, 0, 0)$, respectively. The mean motion of the inner planet is $n = 1$ in these units. Let $(x, y, z)$ denote the position of the test particle, ${v}_{\rm rot}$ denote its velocity in the rotating frame. In CR3BP, the Jacobi energy $E_J$ is an integral of motion and is given by \citep{murray1999solar}
\begin{equation}
\begin{aligned}
    {E_J} &= \frac{1}{2}v_{\rm rot}^2 - \frac{1}{2}{n^2}\left( {{x^2} + {y^2}} \right) - \frac{{1-\mu}}{{{r_{02}}}} - \frac{{\mu}}{{{r_{12}}}}.
\end{aligned}    
\end{equation}
Initially, we have
\begin{equation}
\begin{aligned}
    {{v}_{\rm rot}} &= \left( {\sqrt {\frac{{1-\mu}}{{\gamma^3}}}  - n } \right){\gamma},\\
    r_{02}&=\gamma,\\
    {r_{12}} &= \sqrt {1 + \gamma^2 - 2\gamma \cos \varphi },
\end{aligned}
\end{equation}
where $\gamma=a_2/a_1$ is the ratio of the initial semi-major axes, and $\varphi$ is the angle between the vectors $\vec r_{01}$ and $\vec r_{02}$, with $\vec r_{0i}$ the position vector of planet $i$ relative to the central star. If two planets are set to be co-planar at the beginning, the distance from the test particle to the center of mass is given by
\begin{equation}
    {r_2} = \sqrt {{x^2} + {y^2}}  = \sqrt {\gamma^2 + \mu^2 - 2\gamma{\mu}\cos \varphi }.
\end{equation}
Therefore, $E_J$ can be written as
\begin{equation}
\begin{aligned}\label{eq_ej}
    {E_J} = &\frac{1}{2}{\left( {\sqrt {\frac{{1 - \mu }}{\gamma }}  - \gamma } \right)^2} - \frac{{1 - \mu }}{\gamma } \\& - \frac{1}{2}\left( {{\gamma ^2} + {\mu ^2} - 2\gamma \mu \cos \varphi } \right) \\& - \frac{\mu }{{\sqrt {1 + {\gamma ^2} - 2\gamma \cos \varphi } }},
\end{aligned}
\end{equation}
which is a function of initial angle $\varphi$ for given $\mu$ and $\gamma$. The Jacobi energy $E_J$ achieves its maximum value $E_{J,\max}$ at $\cos \varphi=\gamma/2$, and achieves its minimum value $E_{J,\min}$ at $\cos \varphi=1$ when $\gamma(\gamma^2-1)<1$ and at $\cos \varphi=-1$ when $\gamma(\gamma^2-1)>1$.

The Jacobi energy $E_J$ and the energy of test particle $E_2$ in the inertial frame satisfy the relation 
\begin{equation}
    E_2 = {E_J} + \vec n  \cdot \vec L,
\end{equation}
where $\vec L$ is the angular momentum of the test particle, and $\vec n = (0,0,n)$. After ejection, the energy of the test particle $E_{\rm f}$ is 
\begin{equation}
\begin{aligned}\label{eq_energy}
    {E_{\rm f}} &= \frac{1}{2}v_\infty ^2 = \frac{{1}}{{2\left| {{a_{\rm f}}} \right|}} \\
    &=  {E_J} + \sqrt {\left| {{a_{\rm f}}} \right|\left( {e_{\rm f}^2 - 1} \right)} \cos {i_{\rm f}},
\end{aligned}
\end{equation}
where $a_{\rm f}$, $e_{\rm f}$, and $i_{\rm f}$ are the semi-major axis (scaled by the chosen length unit and therefore dimensionless), eccentricity, and inclination of the test particle measured with respect to the system's center of mass after ejection. Let $q_{\rm f}$ be the the periapsis distance of the hyperbolic trajectory of the test particle after ejection, so ${q_{\rm f}} = \left| {{a_{\rm f}}} \right|\left( {{e_{\rm f}} - 1} \right)$. According to Equation (\ref{eq_energy}), the velocity of the test particle at infinity after ejection is 
\begin{equation}\label{eq_vana}
    {v_\infty } = \sqrt {2\left( {A + \sqrt {{A^2} - B} } \right)},
\end{equation}
where parameters A and B are
\begin{equation}
\begin{aligned}
    A &={E_J} + q_{\rm f}^2{\cos ^2}{i_{\rm f}},\\
    B &= E_J^2 - 2{q_{\rm f}}{\cos ^2}{i_{\rm f}}.
\end{aligned}
\end{equation}
Based on Equation (\ref{eq_vana}), to be ejected, $v_{\infty}$ should be positive, yielding the criterion for ejection  
\begin{equation}\label{eq_cri}
    {q_{\rm f}} > \frac{{E_J^2}}{{2{{\cos }^2}{i_{\rm f}}}}.
\end{equation}
This means that there is a lower limit for the pericenter of the ejected test particle.
At the last close encounter, the separation between the test particle and the planet should be within several Hill radii, so 
\begin{equation}\label{eq_qf}
    q_{\rm f}=1+\alpha r_{\rm H},
\end{equation}
where $r_{\rm H}$ is the dimensionless Hill radius, \[r_{\rm H}={\left( {\frac{{{m_1}}}{{3M}}} \right)^{1/3}},\] and $\alpha$ is typically $1\sim3$ \citep{rodet2024planet}. Therefore, the criterion for ejection represented by Equation (\ref{eq_cri}) can be further written as 
\begin{equation}\label{eq_m}
    \frac{{{m_1}}}{M} > 3{\left( {\frac{{E_J^2}}{{2\alpha {{\cos }^2}{i_{\rm f}}}} - \frac{1}{\alpha }} \right)^3}. 
\end{equation}
Equation (\ref{eq_m}) implies that the planetary mass has a lower bound in order to induce an ejection. In planet–planet scatterings, $i_{\rm f}$ is typically $\lesssim 20^{\circ}$, and $\alpha$ is of order unity, so Equation ($\ref{eq_m}$) provides a practical order-of-magnitude estimate for the minimum planetary mass required for ejection, even if $i_{\rm f}$ and $\alpha$ are not known a priori.

\begin{figure*}
\centering
{\includegraphics[width=0.96\textwidth]{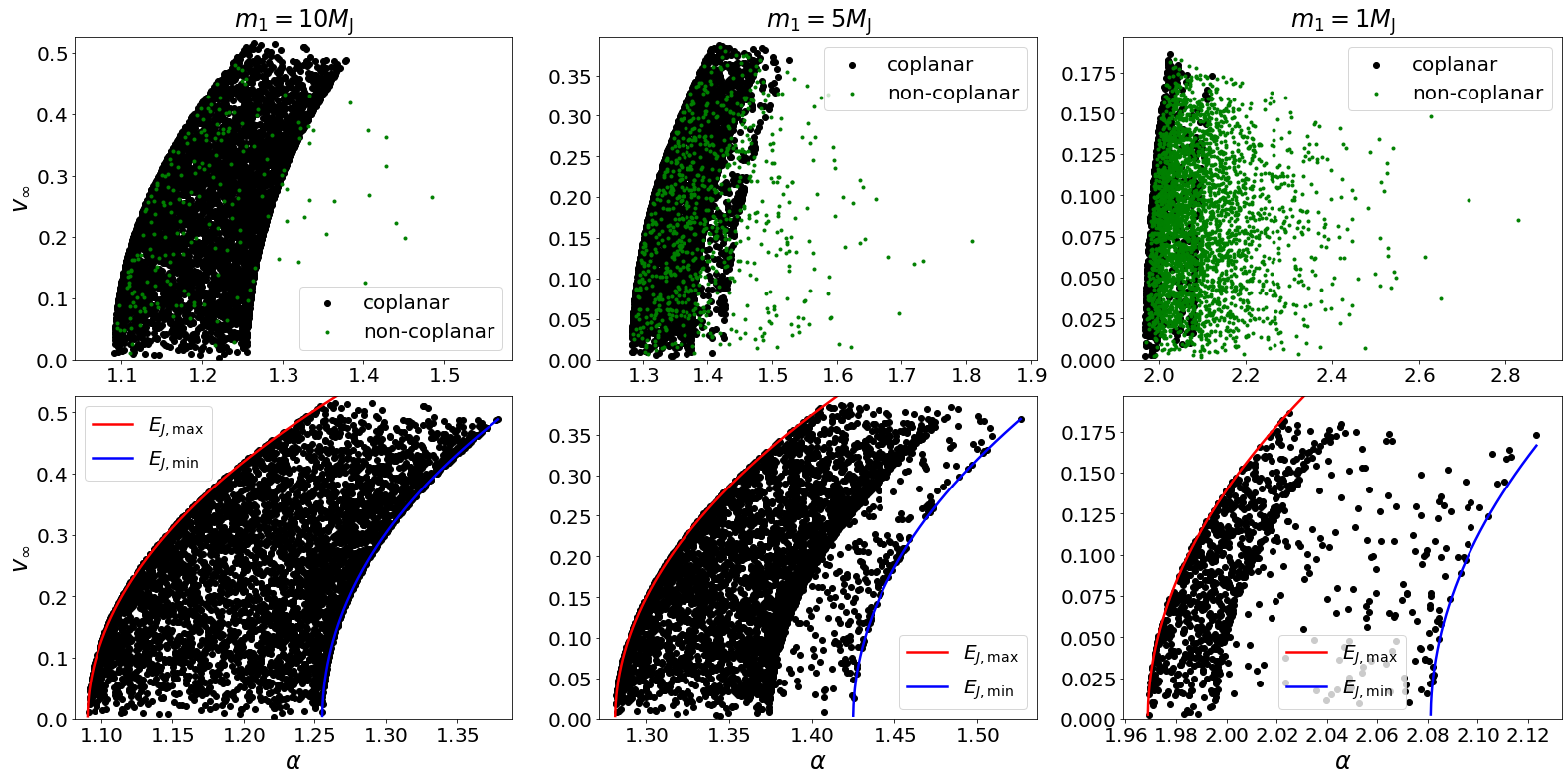}}
\caption{Velocity at infinity vs. parameter $\alpha$ for ejected test particles, where $v_{\infty}$ is in units of $\sqrt{G(M+m_1)/a_1}$, and $\alpha$ is related to the final pericenter distance of the particle. From left to right, the columns correspond to systems with $m_1=10M_{\rm J}$, $5M_{\rm J}$, $1M_{\rm J}$, respectively. In the upper panels, ejection events are classified into two groups: the black dots denote the cases where the scattered particle remains in the orbital plane of the inner planet, while the green dots represent non-coplanar configurations. In the bottom panels, the boundaries of $v_{\infty}$ for the co-planar group are fitted analytically by Equations (\ref{eq_vana}) and (\ref{eq_qf}), where the red and blue lines represent $E_J=E_{J,\max}$ and $E_{J,\min}$ respectively.}
\label{figtpv}
\end{figure*}

\subsection{Numerical test}
We performed simulations for three systems, varying the mass of the inner planet with $m_1=10M_{\rm J}$, $5M_{\rm J}$, $1M_{\rm J}$, respectively. All other parameters are kept identical across the three cases. The outer planet is treated as a test particle with mass $m_2 = 10^{-8}M_{\rm J}$, and its semi-major axis is initially set to $a_2=a_1+2R_{\rm H}$. The planet and test particle are initialized on nearly circular and coplanar orbits, with eccentricities $e_1=e_2=10^{-5}$ and inclinations $i_1=i_2=10^{-3}R_{\rm H}/a_1$. The initial argument of the pericenter, longitude of the ascending node, and mean anomaly are uniformly and randomly chosen in the range $[0, 2\pi]$. We perform 5120 runs for each systems, stopping either when the test particle is ejected or the simulation time exceeds $10^5$ $P_1$, where $P_1$ represents the inner orbital period. We record $v_{\infty}$ after ejections, and compare the simulation outcomes to the analytical result represented by Equation (\ref{eq_vana}).

In the systems with $m_1=10M_{\rm J}$, $5M_{\rm J}$, $1M_{\rm J}$, ejection fractions within the given timescale are $99.2\%$, $98.6\%$, $84.3\%$, respectively. The decrease in ejection efficiency with lower $m_1$ is consistent with the expectation that the mean number of close encounters required for an ejection increases as the planet mass decreases. This relation is approximated by \citep{pu2021strong,li2022long}
\begin{equation}\label{eq_nej}
    \left\langle {{N_{{\rm{ej}}}}} \right\rangle  \simeq {0.06^2}{\left( {\frac{M}{{{m_1}}}} \right)^2}{\left( \frac{m_{12}}{m_1} \right)^4}{\left( {\frac{{{a_2}}}{{{a_1}}}} \right)^{3}}.
\end{equation}
The number of close encounters $N_{\rm{ej}}$ has a broad distribution, but on average, it increases for smaller $m_1$, thereby lengthening the typical timescale required for ejection (see also Section \ref{sv} and Equation (\ref{eq_beta})).

The velocity distributions with respect to the parameter $\alpha$ (see Equation (\ref{eq_qf})) are shown in Figure \ref{figtpv}, where $v_{\infty}$ is in units of $\sqrt{G(M+m_1)/a_1}$ ($\simeq$ 30 km/s for $M=1M_{\odot}$, $a_1=1$ au and $m_1 \ll M$). We determine $\alpha$ numerically by monitoring the pericenter distance of the particle just before ejection. In the upper panels, the black dots represent the outcomes where orbits of the test particle and the inner planet are still co-planar after ejections (a mutual inclination of less than 0.01 is adopted as the criterion for coplanarity), and the green dots represent non-coplanar configurations after ejection. In particular, for the co-planar outcomes, the boundaries of $v_{\infty}$ are analytically reproduced by Equations (\ref{eq_vana}) and (\ref{eq_qf}), where the red line corresponds to $E_J=E_{J,\max}$ and the blue line corresponds to $E_J=E_{J,\min}$. We also see from Figure \ref{figtpv} that (a) the fraction of non-coplanar configuration in the ejection outcomes rises with the decrease of $m_1$, (b)  the maximum value of $v_{\infty}$ is higher with increasing $m_1$, and (c) the density of the black dots in the plots is not uniform. A detailed explanation for point (b) is provided in Section \ref{sv}, while point (c) is discussed further in Appendix \ref{ap1}.

\section{Two-planet scatterings: General cases}\label{sgc}
\subsection{Setup of simulations}\label{sgc_1}
In this section, we consider the general cases, where the outer planet has finite mass, with $m_2 \leq m_1$. In our fiducial simulations, we set $m_1=10M_{\rm J}$, $m_2=1M_{\rm J}$, with mass ratio $m_1/m_2=10$. The outer orbital semi-major axis $a_2$ is given by $a_2-a_1=kR_{\rm H}$, with $k=2$. The inner and outer orbital eccentricities are set to $e_1=e_2 \equiv 0.00001$, the inclinations are  $i_1=i_2 \equiv R_{\rm H}/a_1$, which is roughly $10^{\circ}$ in this case. The initial angles, including argument of the pericenter, longitude of the ascending node, and mean anomaly, are uniformly and randomly chosen in the range $[0, 2\pi]$. We conduct an ensemble of simulations, where each run is terminated either upon the ejection of the outer planet or when the integration time exceeds $10^5\,P_1$. In our simulations, the outer planet is ejected when its distance to the star ($r_{02}$) and to the inner planet ($r_{12}$) exceeds 50 au, and its energy is positive, i.e., $E_2>0$. We have tested different critical distances (e.g., 50 au vs. 1000 au), and found the results to be robust.

We then vary the initial conditions to examine how different parameters influence the outcomes. This analysis is carried out from four perspectives:
\begin{enumerate}
\renewcommand{\labelenumi}{\roman{enumi})}
\item Planetary mass: We vary the mass of the inner planet $m_1$ to $5\,M_{\rm J}$ and $1\,M_{\rm J}$, while keeping the mass ratio fixed at $m_1/m_2 = 10$.
\item Mass ratio: We fix the inner planet mass at $m_1 = 10\,M_{\rm J}$ and vary the mass ratio $m_1/m_2$ to 5, 2, and 1.
\item Initial inclination: We set the initial inclinations of both planets to $i = i_1 = i_2$, and vary $i$ to $0.5\,R_{\rm H}/a_1$, $0.1\,R_{\rm H}/a_1$, and $0.01\,R_{\rm H}/a_1$.
\item Initial planetary separation: We vary the parameter $k$ (defined by $a_2 = a_1 + k R_{\rm H}$) to 2.5, 1.5, and 1.
\end{enumerate}
All other simulation settings remain the same as in the fiducial model.

As expected, the outer planet is ejected for systems with $m_2<m_1$. For systems with $m_2=m_1$, it is also possible for the inner planet to be ejected. In general, a more massive inner planet or a larger mass ratio leads to a higher fraction of ejection in the given timescale. We discuss the minimum separations between the two planets prior to ejection in Section~\ref{sr}, and analyze the ejection velocity distribution in Section~\ref{sv}. In all figures, the results from the fiducial simulations are shown as black curves for direct comparison.

\subsection{Results: Minimum planet-plant separation prior to ejection}\label{sr}
The minimum separation $r_{12,\min}$ between the two planets prior to ejection plays a crucial role in determining the survivability of moons. As noted before (see Equation (\ref{eq_r2m})), if $r_{12,\min}$ is less than the moon’s orbital radius around its host planet (for $m_1 \sim m_2$), the moon is likely to be disrupted through collision with the planet, capture by the star, or ejection from the system \citep{hong2018innocent}. Conversely, if the moon’s orbital radius is smaller than $r_{12,\min}$, it may avoid strong perturbations and remain bound to its host.

To quantify this, we analyze the statistical distribution of $r_{12,\min}$ for our simulations. We present both the cumulative distribution function (CDF) and the complementary cumulative distribution function (CCDF = 1 $-$ CDF) of $r_{12,\min}$. The CDF gives the probability that a moon with orbital radius $\sim r_{12,\min}$ would be disrupted (see Equation (\ref{eq_r2m}) for more precise criterion). The CCDF represents the survival probability with orbital radius $\sim r_{12,\min}$.

\subsubsection{Systems of different inner planetary masses}
\begin{figure*}
\centering
{\includegraphics[width=0.96\textwidth]{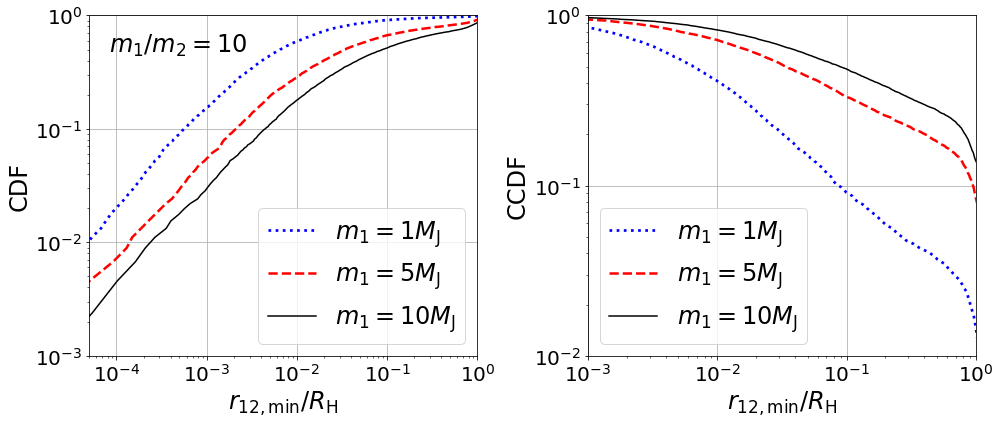}}
\caption{The cumulative distribution (CDF) and complementary cumulative distribution (CCDF) of minimum separation between the planets, $r_{12,\min}$ (scaled by the mutual Hill radius $R_{\rm H}$) in systems of different planetary masses. The black, red and blue lines represent systems with $m_1=10M_{\rm J}$, $5M_{\rm J}$, $1M_{\rm J}$, respectively, with the stellar mass $M=1M_{\odot}$. Other parameters are the same for each system, with mass ratio $m_1/m_2=10$, initial semi-major axes $a_1=1$au, $a_2=a_1+2R_{\rm H}$, eccentricity $e_1=e_2=10^{-5}$, inclination $i_1=i_2=R_{\rm H}/a_1$. Initial argument of the pericenter, longitude of the ascending node, and mean anomaly are uniformly and randomly chosen in the range $[0, 2\pi]$.} 
\label{figrm1}
\end{figure*}

Figure \ref{figrm1} shows CDF and CCDF of the minimum separation $r_{12,\min}$ (scaled by mutual Hill radius $R_{\rm H}$) for different inner planetary masses, with mass ratio $m_1/m_2$ fixed. The black, red and blue curves represent $m_1=10M_{\rm J}$, $m_1=5M_{\rm J}$, and $m_1=1M_{\rm J}$ respectively, and the medians of $r_{12,\min}/R_{\rm H}$ corresponding to these curves are 0.089, 0.034, 0.0066, respectively.

At any fixed $r_{12,\min}/R_{\rm H}$, the CDF value is lower for larger $m_1$, indicating that lower probability for such close encounters. This suggests that as $m_1$ increases, the typical minimum encounter distance before ejection tends to be larger. This trend can be understood by the fact that a more massive planet exerts stronger gravitational force on the other planet, but a less massive planet should get closer to the other planet in order to cause an ejection. The trend in the CDF indicates that moons orbiting smaller planets (e.g., $m_1 = 1\,M_{\rm J}$) are more likely to be disrupted, as close encounters occur more frequently. In contrast, systems with more massive planets (e.g., $m_1 = 10\,M_{\rm J}$) experience fewer deep encounters, allowing moons to survive more easily. 

\subsubsection{Systems of different planet mass ratios}
\begin{figure*}
\centering
{\includegraphics[width=0.96\textwidth]{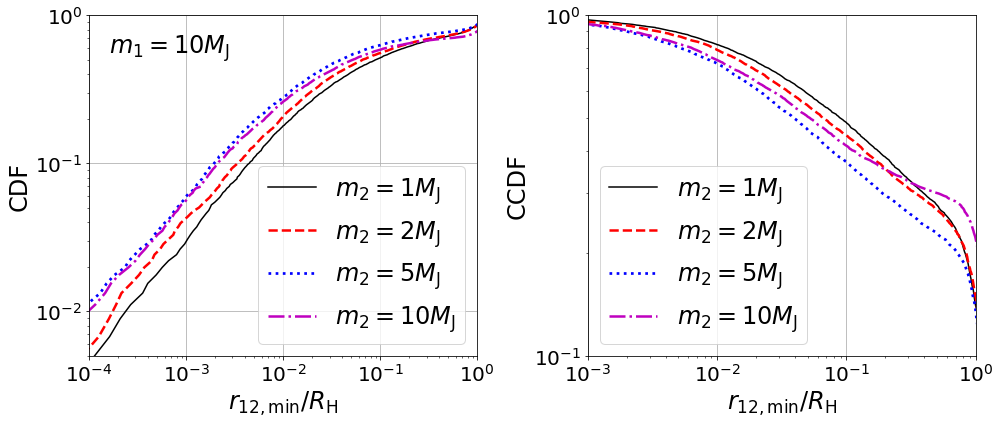}}
\caption{Same as Figure \ref{figrm1}, but in systems of different planetary mass ratios and with $m_1$ fixed at 10$M_{\rm J}$. The black, red, blue, and magenta lines represent systems of $m_2=1M_{\rm J}$, $2M_{\rm J}$, $5M_{\rm J}$, $10M_{\rm J}$, respectively, corresponding to mass ratio of 10, 5, 2, 1.}
\label{figrmr}
\end{figure*}

Figure \ref{figrmr} shows the CDF and CCDF of $r_{12,\min}/R_{\rm H}$ for systems with $m_2=1M_{\rm J}$, 2$M_{\rm J}$, $5M_{\rm J}$, and $10M_{\rm J}$, respectively, with $m_1$ fixed at $10M_{\rm J}$. The corresponding medians of $r_{\min}/R_{\rm H}$ are 0.089, 0.065, 0.039, and 0.052.

Except for the case of $m_2=m_1=10M_{\rm J}$, where both planets may be ejected, in general systems with smaller $m_1/m_2$ tend to experience closer encounters before ejection, as reflected in the CDF shift toward smaller $r_{12,\min}/R_{\rm H}$ for increasing $m_2$. Nevertheless, a comparison between Figure \ref{figrm1} and Figure \ref{figrmr} suggests that mass ratio has less influence on the survival rate of moons compared to planet mass $m_1$.

\begin{figure*}
\centering
{\includegraphics[width=0.96\textwidth]{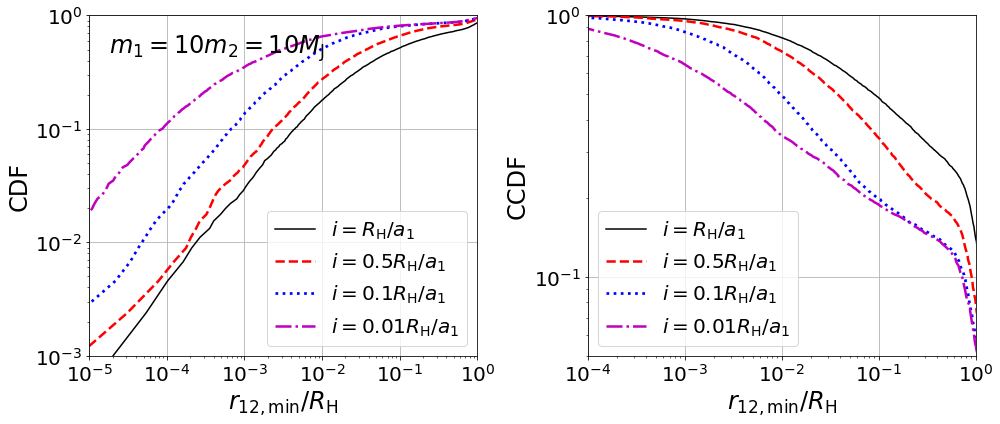}}
\caption{Same as Figure \ref{figrm1}, but in systems of different initial inclinations and with $m_1$ fixed at $10M_{\rm J}$, $m_1/m_2$ fixed at 10. Initial inclinations of $i=i_1=i_2=R_{\rm H}/a_1$, $0.5R_{\rm H}/a_1$, $0.1R_{\rm H}/a_1$, $0.01R_{\rm H}/a_1$ are denoted by black, red, blue, and magenta lines, respectively.}
\label{figri}
\end{figure*}

\begin{figure*}
\centering
{\includegraphics[width=0.96\textwidth]{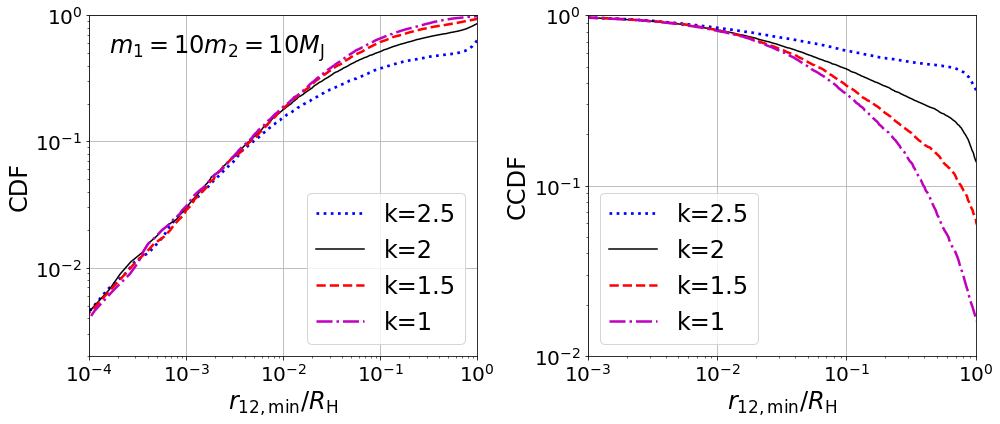}}
\caption{Same as Figure \ref{figrm1}, but in systems of different initial planetary separations and with $m_1$ fixed at $10M_{\rm J}$, $m_1/m_2$ fixed at 10. Initial planetary separation $a_2-a_1 =kR_{\rm H}$, with $k=2.5$, 2, 1.5, 1 are denoted by blue, black, red, and magenta lines, respectively.}
\label{figrk}
\end{figure*}

\begin{figure}
{\includegraphics[width=0.49\textwidth]{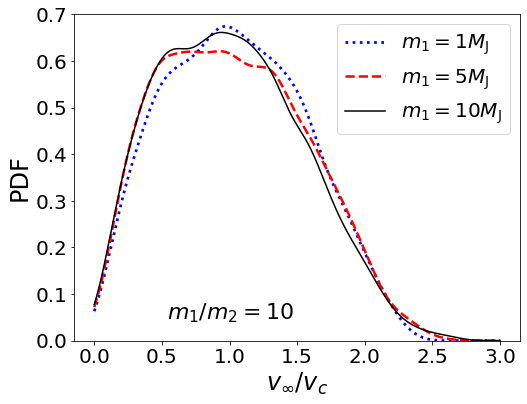}}
\caption{Probability distribution of $v_{\infty}/v_c$ in the systems of different planetary masses, where $v_{\infty}$ is the velocity of the ejected planet at infinity. Initial settings are the same as those in Figure \ref{figrm1}.}
\label{figvm1}
\end{figure}

\begin{figure}
{\includegraphics[width=0.49\textwidth]{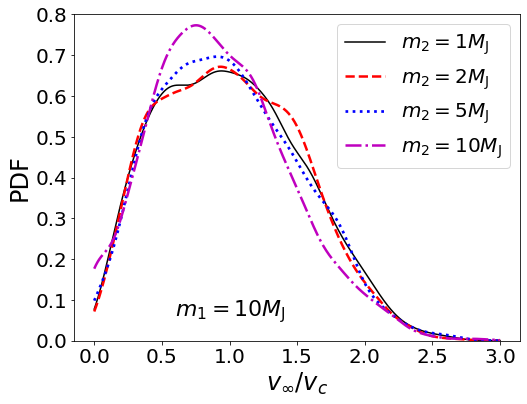}}
\caption{Same as Figure \ref{figvm1}, but in the systems of different planetary mass ratios. Initial settings are the same as those in Figure \ref{figrmr}.}
\label{figvmr}
\end{figure}

\begin{figure}
{\includegraphics[width=0.49\textwidth]{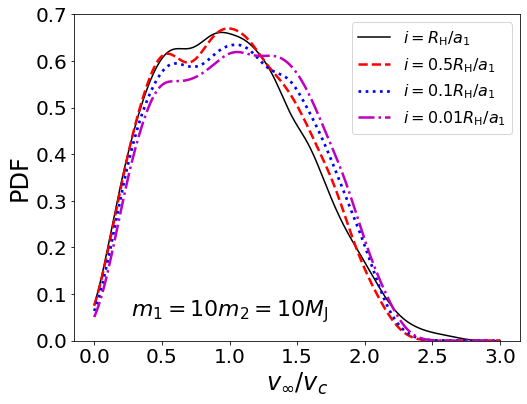}}
\caption{Same as Figure \ref{figvm1}, but in the systems of different initial inclinations. Initial settings are the same as those in Figure \ref{figri}.}
\label{figvi}
\end{figure}

\begin{figure}
{\includegraphics[width=0.49\textwidth]{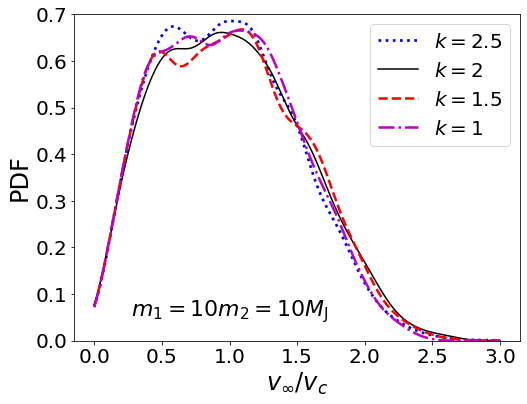}}
\caption{Same as Figure \ref{figvm1}, but in the systems of different initial planetary separations. Initial settings are the same as those in Figure \ref{figrk}.}
\label{figvk}
\end{figure}

\subsubsection{Systems of different initial inclinations}
Figure \ref{figri} shows the CDF and CCDF of $r_{12,\min}$ for systems with initial orbital inclinations $i_1=i_2=R_{\rm H}/a_1$ (black), $0.5R_{\rm H}/a_1$ (red), $0.1R_{\rm H}/a_1$ (blue), and $0.01R_{\rm H}/a_1$ (magenta). The medians of $r_{12,\min}/R_{\rm H}$ for these curves are 0.089, 0.038, 0.0096, 0.0030, respectively. These results suggest that systems with smaller inclinations tend to have closer planet-planet encounters, making moons less likely to survive.

\subsubsection{Systems of different initial planetary separations}
Figure \ref{figrk} shows the CDF and CCDF of $r_{12,\min}/R_{\rm H}$ for different values of initial $k=(a_2-a_1)/R_{\rm H}$. The medians of $r_{12,\min}/R_{\rm H}$ for the $k=2.5$ (blue), 2 (black), 1.5 (red), and 1 (magenta) curves are 0.60, 0.089, 0.060, 0.051, respectively. We see that for $r_{12,\min}/R_{\rm H} \lesssim 10^{-2}$, the initial separation of the two planets has little impact on the CDF. For $r_{12,\min}/R_{\rm H} \gtrsim 10^{-2}$, the CDFs diverge for different initial separations. The CCDF curve decreases rapidly for a smaller $k$. Therefore, if a moon orbits its planet at radius less than $10^{-2}R_{\rm H}$, different initial separations lead to similar survival possibility of the moon, which is at least 70\%. If the orbital radius of the moon is larger than $10^{-2}R_{\rm H}$, its survival rate after planet-planet ejection increases with the initial planet separation.

\subsection{Results: Velocities of FFPs}\label{sv}
Now we study the velocity distribution of FFPs.We record the velocity $v_{\infty}$ of the ejected planet at infinity using Equation (\ref{eq_vinf}), and obtain the probability density function (PDF) under different initial settings by using a Gaussian kernel density estimator.

Figures \ref{figvm1}, \ref{figvmr}, \ref{figvi}, and \ref{figvk} show the PDFs of $v_{\infty}$ of the ejected planet $m_2$. The corresponding initial settings are the same as those in Figures \ref{figrm1}, \ref{figrmr}, \ref{figri}, and \ref{figrk}, respectively. The velocity distributions are all scaled by a critical value $v_c$, given by
\begin{equation}\label{eq_vc}
    {v_c} = {\left( {\frac{{G{m_1}}}{{0.12{a_2}}}} \right)^{1/2}}\left( {\frac{{{m_1}}}{m_{12}}} \right){\left( {\frac{{{a_1}}}{{{a_2}}}} \right)^{1/4}}.
\end{equation}
After being scaled by $v_c$, the velocity distributions in Figures \ref{figvm1}, \ref{figvmr}, \ref{figvi}, and \ref{figvk} are similar, with $v_{\infty}/v_c$ ranging from 0 to 2.5, and with a peak around unity. In addition, we found that the initial inclinations have little influence on the velocities of the ejected planets.

The critical value $v_c$ can be understood as follows. During a close encounter, the two planets exchange energy. The outer planet may gain or loss energy by $\Delta E \sim \beta(Gm_1m_2/a_1)$, where $\beta$ is a dimensionless number of order a few. By cumulatively gaining energy from the inner planet, the outer one is finally ejected when its energy becomes positive. The mean number of close encounter is $\left\langle {{N_{{\rm{ej}}}}} \right\rangle \sim (GMm_2/2a_1)^2/\Delta E^2$ \citep{pu2021strong,li2022long}. By combining Equation (\ref{eq_nej}), we have
\begin{equation}\label{eq_beta}
    \beta  \simeq \frac{1}{{0.12}}{\left( {\frac{{{m_1}}}{{m_{12}}}} \right)^2}{\left( {\frac{{{a_1}}}{{{a_2}}}} \right)^{3/2}}.
\end{equation}
Therefore, the velocity of FFPs can be estimated as $v_c \sim \sqrt{\Delta E/m_2} \sim \sqrt{\beta Gm_1/a_1}$. By applying Equation (\ref{eq_beta}), we obtain the estimated velocity of FFPs represented by Equation (\ref{eq_vc}). Note that Equation (\ref{eq_vc}), $v_c \simeq 3(m_1/m_{12})\sqrt{Gm_1/a_2}$ ($\simeq$ 3 km/s for $m_1=1M_{\rm J}$, $a_2=1$ au, and $m_2\ll m_1$), is independent of the stellar mass $M$; it applies only to the mass ratio range considered in this section $(M/m_1=10^2 \sim 10^3)$. For very large $M/m_1$, no ejection is possible (see Section \ref{stp}), and Equation (\ref{eq_vc}) obviously breaks down. The ejection velocity in our results is consistent with \citet{bhaskar2025properties} and \citet{Zhai_2025}, where FFPs are also produced by planet-planet scatterings but with very different initial settings.

\subsection{Finite sizes of the planets}
In this paper, we focus on the ejection velocity distribution of FFPs in the scale-free setups, where the planets are considered as point masses. Here we discuss the influence of the finite sizes of the planets.

\begin{figure}
{\includegraphics[width=0.49\textwidth]{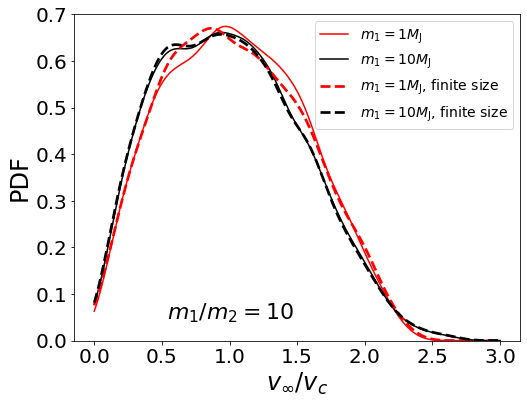}}
\caption{Probability distribution of $v_{\infty}/v_c$ in the systems of different planetary masses, where $v_{\infty}$ is the velocity of the ejected planet at infinity. The solid lines are for planets with zero radius, and the dashed lines for planets with radius equal to $R_{\rm J}$. The initial settings are the same as in Figure \ref{figrm1}.}
\label{figsize}
\end{figure}

When considering the sizes of the planets, the outcomes of the dynamical instability include collisions. We set the radius of each planet to the Jupiter radius $R_{\rm J}$. If the minimum separation between two planets is smaller than $2R_{\rm J}$, we regard it as a planet-planet collision. Obviously our calculation of the survival of the moons is not affected by the finite planet sizes. To determine the ejection velocity of FFPs, we exclude the collision outcomes by removing the cases where $r_{12,\min} \leq 2R_{\rm J}$ and show the ejection velocity distribution for the `real' ejection outcomes in Figure \ref{figsize}. We only show the fiducial runs and the one with the largest fraction of collisions among all the systems. In these two runs, the ratio of collisions and ejections is 0.127 and 1.707, respectively. We see that the distributions are similar, with a characteristic ejection velocity $v_c$ given by Equation (\ref{eq_vc}). Thus, considering the finite sizes of the planets does not change our conclusions.

\section{Three-planet scatterings}\label{s3p}
In Section \ref{sgc}, we have studied planet-planet scatterings in two-planet systems. But planetary systems may harbor more than two planets. In this section, we consider three-planet scatterings to produce FFPs, and also study the survival rate of hypothetical moons around the ejected planet and ejection velocity of FFPs. 

\subsection{Setup of simulations}
In the simulations, we set the three planets in a close configuration, with
\begin{equation}
    a_{i+1}-a_{i}=kR_{\rm H},
\end{equation}
where $R_{\rm H}$ is the mutual Hill radius,
\begin{equation}\label{eq_3rh}
    {R_{\rm H}}{\rm{ = }}\frac{{{a_i} + {a_{i+1}}}}{2}{\left( {\frac{{{m_i} + {m_{i+1}}}}{{3M}}} \right)^{1/3}}.
\end{equation}
The mass of the star is set to $M=1M_{\odot}$, the semi-major axis of the innermost planet $a_1$ is set to 1 au. In our fudicial runs, the planetary masses are set to $m_1=5M_{\rm J}$, $m_2=2M_{\rm J}$, and $m_3=2M_{\rm J}$ (Hereafter the fiducial three-planet system is denoted as `522'). Each planet has an eccentricity $10^{-5}$ and inclination $0.1R_{{\rm H}ij}/a_1$ where $R_{{\rm H}{ij}}$ is the mutual Hill radius of planet $i$ and planet $j$. The parameter $k$ is set to $k=3$. The argument of pericenter, longitude of ascending node, and mean anomaly of each planet are randomly and uniformly chosen in the range $[0,2\pi]$. We perform over 5000 $N$-body simulations, and stop either when one planet is ejected or the simulation time exceeds $10^{5}P_1$. 

As in Section \ref{sgc_1}, the ejection criterion for planet $i$ is $r_{0i}>50$ au, $r_{ij}>50$ au, $r_{ik}>50$ au and $E_{i}>0$, where $E_i$ is the energy of the ejected planet, given by
\begin{equation}
    {E_i} = \frac{1}{2}v_i^2 - \frac{{GM}}{{{r_{0i}}}} - \frac{{G{m_j}}}{{{r_{ij}}}} - \frac{G{m_k}}{r_{ik}},
\end{equation}
and $r_{0i}$ is the separation between the ejected planet and the star, $r_{ij}$ and $r_{ik}$ are the separations between the ejected planet and the other two planets $j$ and $k$. Similar to the two-planet scattering case, before ejection we record the minimum separations of one planet with respect to the other two to estimate the survival rate of moons around the ejected planet.

In addition to the fiducial runs, we also cary out the following sets of simulations:
\begin{enumerate}
  \renewcommand{\labelenumi}{\roman{enumi})} 
  \item Planet masses $m_1=m_3=2M_{\rm J}, m_2=5M_{\rm J}$ (hereafter denoted as '252'), and $m_1=m_2=2M_{\rm J}, m_3=5M_{\rm J}$ (hereafter denoted as '225');
  \item With $m_1$ fixed at $5M_{\rm J}$, the other two planet masses $m_2=2M_{\rm J},m_3=1M_{\rm J}$ (hereafter denoted as `521') and $m_2=1M_{\rm J},m_3=1M_{\rm J}$ (hereafter denoted as `511');
  \item Separation parameter $k=3.6$ in `522'.
\end{enumerate}

In the above simulations, over 99.4\% of the outcomes are planet ejection, and the rest still remains bound for the given running times. In the cases of ejection, the probability of the most massive planet being ejected is negligible ($<0.5$\%). For systems `522', `252', `225', and `511', the ejection probabilities of two less massive planets are comparable. For system `521', the ejection probability of $m_3$ is 2.27 times that of $m_2$. 

\subsection{Results: Minimum planetary separations prior to ejection}\label{s3pr}

\begin{figure*}
{\includegraphics[width=0.99\textwidth]{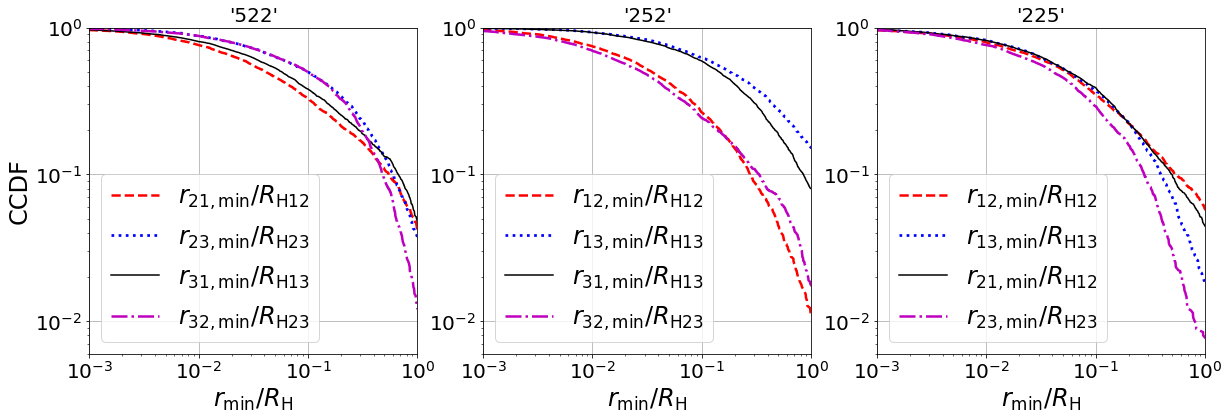}}
\caption{The CCDFs of the minimum separations $r_{ij,{\rm min}}$ between the ejected planet $i$ and another bound planet $j$ (scaled by their initial mutual Hill radius) for three different simulation sets as labeled (e.g., `522' refers to $m_1=5M_{\rm J}$, $m_2=m_3=2M_{\rm J}$).}
\label{fig3pro}
\end{figure*}

\begin{figure*}
\centering
{\includegraphics[width=0.99\textwidth]{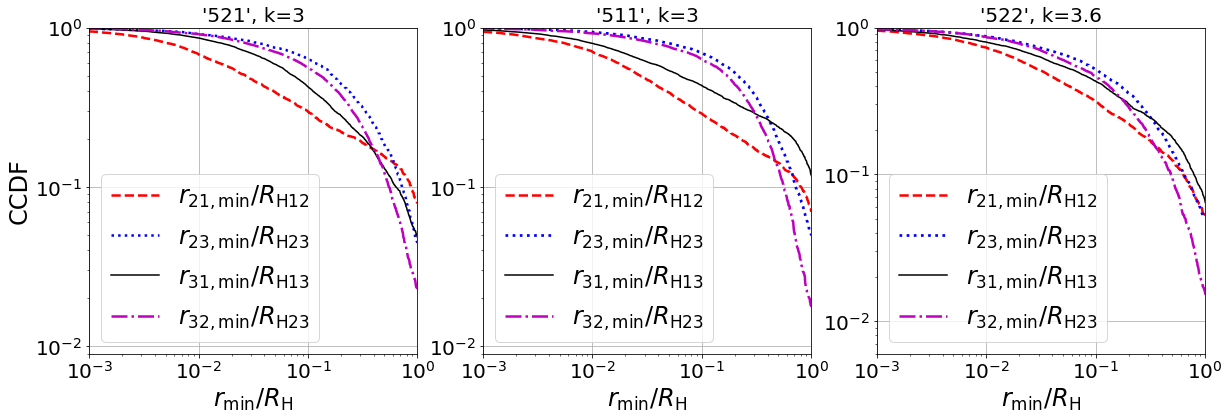}}
\caption{Same as Figure \ref{fig3pro}, but for different planetary masses and initial planetary separations.}
\label{fig3prm}
\end{figure*}

Figures \ref{fig3pro} and \ref{fig3prm} show CCDFs of the minimum separations between every two planets in our simulation sets, where $r_{ij,{\rm min}}$ is the minimum separation between the ejected planet $i$ and another planet $j$ which is still bound in the system. The minimum separations shown in the CCDFs are scaled by the initial mutual Hill radius of planets $i$ and $j$ (see Equation (\ref{eq_3rh})). Comparing the red lines to the blue lines or comparing the black lines to the magenta ones, we see that for the same ejected planet, the percentage of closer encounters between the ejected planet and the most massive planet is higher than that between the ejected planet and the other less massive planet. In other words, the survival of moons around the ejected planet mainly depends on the separation of close encounters with the most massive planet in the system. Also, considering semi-major axes of moons that are similar to those in our Solar system ($\lesssim 0.1R_{\rm H}$), the orderings of planets have a small influence on survival probability of hypothetical moons, and the two less massive planets have similar probabilities of keeping their moons, as shown in Figure \ref{fig3pro}. In addition, according to the red lines and the black lines in Figure \ref{fig3prm}, the outer planet has a higher probability to keep its moons compared to the middle planet. Nevertheless, for $r_{ij,\min}<0.1R_{\rm H}$, the CCDFs corresponding to the minimum separation between the ejected planet and the most massive planets in different systems have no significant difference, and the survival rate of potential moons with orbital radii less than $0.1 R_{\rm H}$ are $\sim 30\%-40\%$. Also, the initial planetary separation has little influence on the moon's survivability.

\subsection{Results: Velocities of FFPs}\label{s3pv}
\begin{figure}
{\includegraphics[width=0.49\textwidth]{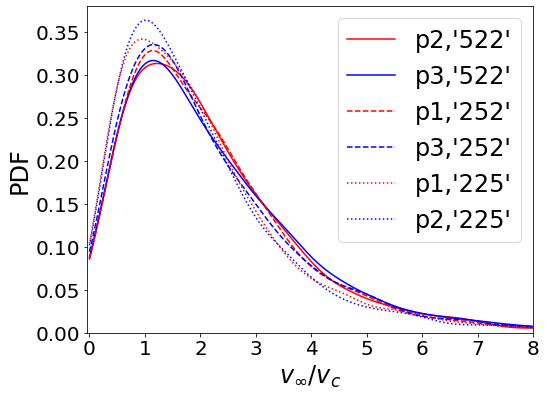}}
\caption{The PDF of velocity of the ejected planets in different systems. The ejection velocity $v_{\infty}$ is scaled by a typical value $v_c$ which is represented by Equation (\ref{eq_3pvc}). The labels 'p$i$' represent the ejected planet.}
\label{fig3pvo}
\end{figure}
\begin{figure}
{\includegraphics[width=0.49\textwidth]{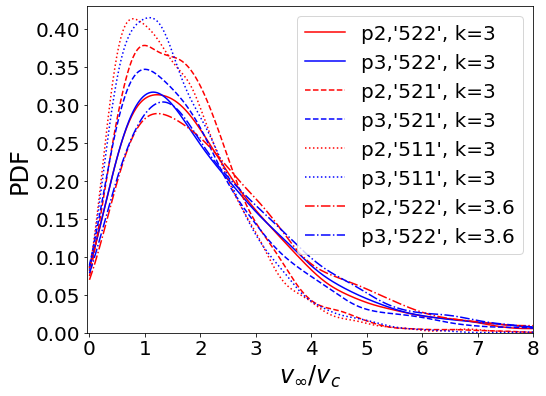}}
\caption{Same as Figure \ref{fig3pvo}, but for different planetary masses and initial planetary separations.}
\label{fig3pvmk}
\end{figure}

The PDFs of the ejection velocity $v_{\infty}$ for our different simulation runs are shown in Figures~\ref{fig3pvo} and~\ref{fig3pvmk}, where the symbol `p$i$' in the legends denotes that planet $i$ is the one being ejected. The velocity $v_{\infty}$ is normalized by a characteristic scale $v_c$, defined similarly to Equation~(\ref{eq_vc}) as:
\begin{equation}\label{eq_3pvc}
    {v_c} = {\left( {\frac{{G{m_p}}}{{0.12{a_3}}}} \right)^{1/2}}\left( {\frac{{{m_p}}}{{{m_p} + {m_i}}}} \right){\left( {\frac{{{a_1}}}{{{a_3}}}} \right)^{1/4}},
\end{equation}
where $m_p$ represents the largest planetary mass in the system, and $m_i$ represents the mass of the ejected planet $i$. We see from Figures \ref{fig3pvo} and \ref{fig3pvmk} that the velocity distributions all have a peak around $v_{\infty}/v_c \simeq 1$, demonstrating that $v_c$ serves as a reliable estimator for the typical ejection velocity in three-planet systems. One difference, however, is that for two-planet systems, the PDF extends to $v_{\infty}/v_c \simeq 2.5$, but for three-planet systems, the PDF is broader and extends to $v_{\infty}/v_c \simeq 5 - 8$.

\section{Ejection Velocities of Circumbinary Planets}\label{scbp}
\begin{figure}
{\includegraphics[width=0.49\textwidth]{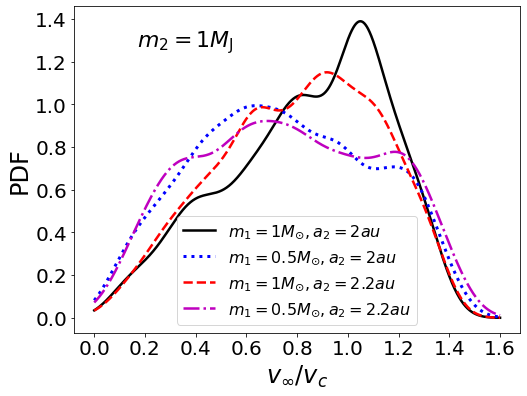}}
\caption{The PDF of ejection velocity of the circumbinary planets in different systems. The ejection velocity $v_{\infty}$ is scaled by the characteristic value $v_c$ given by Equation (\ref{eq_vcbp}).}
\label{figcbp}
\end{figure}

To compare with planet-planet scatterings, we study the ejection velocity of the circumbinary planets in this section (also see \cite{coleman2024properties}). We assume initially a planet orbits a binary in an unstable configuration (see \citealp{holman1999long} and \citealp{georgakarakos2024empirical} for specific unstable criteria), with the semi-major axis of the binary $a_1=1$ au, the semi-major axis of the planet with respect to the binary's center of mass $a_2=2$ au or 2.2 au. One component of the binary is set to  $m_0=1M_{\odot}$, and the other one is set to $m_1=1M_{\odot}$ or $0.5M_{\odot}$. The mass of the circumbinary planet is set to 1 $M_{\rm J}$. The eccentricities of the inner and outer orbits are set to $10^{-5}$, and the inclinations are both set to 0. The mean anomaly of each orbit are randomly and uniformly chosen in the range $[0,2\pi]$. We perform over 5000 $N$-body simulations, and stop either when the circumbinary planet is ejected or the simulation time exceeds $10^{5}P_1$, where $P_1$ is the orbital period of the inner binary. The ejection velocity distributions are shown in Figure \ref{figcbp}, scaled by $v_c$ which is defined similarly to Equations (\ref{eq_vc}) and (\ref{eq_3pvc}):
\begin{equation}\label{eq_vcbp}
     {v_c} = {\left( {\frac{{G{m_1}}}{{0.12{a_2}}}} \right)^{1/2}}\left( \frac{m_0}{m_0+m_1}\right){\left( {\frac{{{a_1}}}{{{a_2}}}} \right)^{1/4}}.
\end{equation}
In Equation (\ref{eq_vcbp}), $v_c$ is modified by a mass factor $m_0/(m_0+m_1)$, reflecting the fact that the planet is scattered by the central binary composed of both $m_0$ and $m_1$. We see from Figure \ref{figcbp} and Equation (\ref{eq_vcbp}) that the ejection velocity of circumbinary planet is typically one order of magnitude higher than the velocity of FFP produced by planet-planet scatterings due to the increase of $m_1$.

\section{Conclusion and discussion}\label{scon}
FFPs can form through various channels, among which planet–planet scatterings is considered one of the most plausible, as it commonly occurs in dynamically unstable multi-planet systems. In such systems, close encounters can lead to a range of outcomes, including planetary collisions, engulfment by the central star, or ejection from the system. In this work, we focus on FFPs formed via planet–planet scatterings, examining their ejection velocities and assessing the likelihood that moons—if originally present—remain gravitationally bound after ejection. We analyze both two-planet and three-planet systems. To isolate the key dynamics, we do not explicitly include moons in our simulations; instead, we estimate the survivability of potential moons based on the minimum planet–planet separations recorded prior to ejection. To obtain robust results that can be scaled to different systems, we treat the stars and planets as point masses, thereby excluding physical collisions from the outcomes.  

In the framework of CR3BP, we analyze the ejection velocity distributions for different values of the planet mass $m_1$, as shown in Figure~\ref{figtpv}. The analytical boundaries of the ejection velocities, derived from Jacobi energy conservation (Equation (\ref{eq_vana})), match well with the simulation results. Our analysis reveals the existence of a minimum planetary mass $m_1$ required to produce ejection events.

We explore systems in which both planets have finite masses using $N$-body simulations. Our results for the distributions of the minimum planet-planet separations prior to ejections are displayed in various figures, which can be used to determine the survivability of moons with different orbital radii. For example, in our fiducial simulations (with $m_1/M \simeq 10^{-2}$, $m_1/m_2 = 10$), we find that the probability of the minimum planet–planet separation satisfying $r_{\min} > 0.1 R_{\rm H}$ is 48\%, and for $r_{\min} > 0.01 R_{\rm H}$, it is 82\% (see Figures \ref{figrm1} -- \ref{figrk}). Given that the regular moons in the Solar System typically orbit at $\sim 10^{-2} R_{\rm H}$ from their planet hosts, this suggests that a significant fraction of moons can survive planet-planet scatterings and many free-floating giant planets may have retained their moons. We also find that the moon survival probability somewhat decreases as the planetary mass $m_1$ and the mass ratio $m_1/m_2$ reduced. Note that although physical collisions are excluded by treating the star and planets as point masses, including their finite sizes would not affect the moon survival statistics as long as the planetary radius is much less than $r_{\min}$. 

We have determined the velocity distributions of the ejected planets for various system parameters (see Figures \ref{figvm1} -- \ref{figvk}). We find that the peak of the distribution can be specified by an analytical expression (Equation (\ref{eq_vc})). Our results show that a more massive planet in the system leads to higher ejection velocities. Incorporating the physical radii of the planets does not change our results, reinforcing the robustness of our findings.

For three-planet systems, despite the increased dynamical complexity, the peak of the velocity distribution of the ejected planet can be well approximated by a similar analytical expression $v_c$ (see Equation \ref{eq_3pvc}), although the distribution can extend to (5 -- 8)$v_c$. For the circumbinary planetary systems, the ejection velocity of FFPs is typically one order of magnitude higher than the ejection velocity of planets which are produced by planet-planet scatterings (see Equation \ref{eq_vcbp}).

We note that the velocity of a FFP is the vector sum of the ejection velocity and the velocity of its original parent star, the latter has a magnitude of $v_{*} \sim 10$ km/s, which is the typical velocity dispersion of the young stars. Equation (\ref{eq_vc}) or (\ref{eq_3pvc}) shows that for a small planet ejected by a giant planet (with mass $m_p$ and semi-major axis $a_p$), the typical ejection velocity is 
\begin{equation}
    v_c  \simeq (3 {\rm km/s})\left(\frac{m_p}{M_{\rm J}}\right) ^{1/2}\left(\frac{a_p}{1 \rm{au}}\right) ^{-1/2}.
\end{equation}
This is smaller than $v_{*}$ unless $m_p \gtrsim M_{\rm J}$ and/or $a_p \lesssim 1$ au. For example, a super Earth (SE) ejected by a warm Jupiter (WJ, $a_p \sim 0.1$ au) would have an ejection velocity comparable to or larger than $v_*$. Of course, since WJs have Safronov number (the squared ratio of the escape velocity from the planetary surface to the planet's orbital velocity) of order unity, SE-WJ scatterings would lead to both planet ejection and collision (e.g., \citealp{anderson2020situ}). In practice, constraining the ejection velocities from the ``observed" velocities of FFPs maybe challenging, and will likely require a large sample.

Overall, our results demonstrate that FFPs produced via planet–planet scatterings have a significant probability of retaining their moons. Moreover, the ejection velocity of such planets serves as a diagnostic of their dynamical origin and the properties of their parent systems. Our results may help unravel the dynamical history of FFPs using observations and offer promising prospects for the future detection of exomoons around FFPs.

\begin{acknowledgments}
X.H. thanks Jiaru Li and Fangyuan Yu for useful discussions.
\end{acknowledgments}

\appendix
\twocolumngrid
\section{Density of ejection velocity distribution in CR3BP}
\label{ap1}
\begin{figure}
\centering
{\includegraphics[width=0.49\textwidth]{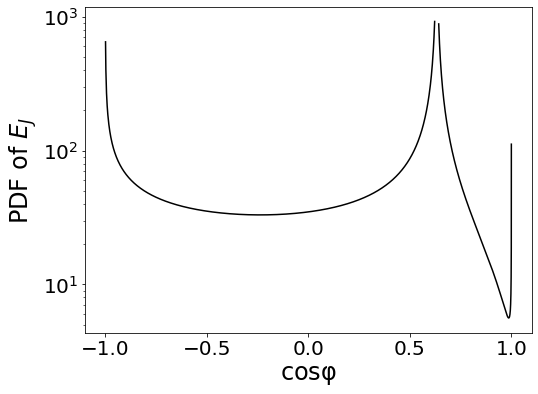}}
\caption{Probability distribution of $E_J$ as a function of $\cos \varphi$.}
\label{figEJ}
\end{figure}

The density of the black dots shown in Figure \ref{figtpv} is not uniformly distributed and 
is different for the three systems. The reasons are as follows. The initial orbital phases of the inner and outer planets are randomly chosen in $[0,2\pi]$, so the probability density function (PDF) of $\cos \varphi$ is 
\begin{equation}
    \left| {\frac{{{\rm d}P}}{{{\rm d}\cos\varphi }}} \right| = \frac{1}{{2\pi }}\left| {\frac{1}{{\sin \varphi }}} \right|.
\end{equation}
The PDF of $E_J$ (Equation (\ref{eq_ej})) is then
\begin{equation}
\begin{aligned}\label{eq_dej}
    \left| {\frac{{{\rm d}P}}{{{\rm d}{E_J}}}} \right| &= \left| {\frac{{{\rm d}P}}{{{\rm d}\cos \varphi }}\frac{{{\rm d}\cos \varphi }}{{{\rm d}{E_J}}}} \right| \\ &= \frac{1}{{2\pi }}{\left[ {\left| {\sin \varphi \left( {\gamma \mu  - \frac{{\gamma \mu }}{{{{\left( {1 + {\gamma ^2} - 2\gamma \cos \varphi } \right)}^{3/2}}}}} \right)} \right|} \right]^{ - 1}}.
\end{aligned}
\end{equation}
As an example, for $m_1=5M_J$, the PDF of $E_J$ as a function of $\cos \varphi$ is shown in Figure \ref{figEJ}. When $\cos \varphi=\pm 1$ or $\cos \varphi=\gamma/2$, the PDF is infinite. In addition, the PDF has a minimum of $\cos \varphi = \cos \varphi_c$, with $\cos \varphi_c \in (\gamma/2,1)$. Therefore, black dots in Figure \ref{figtpv} is very dense around $E_{J,\max}$ and $E_{J}(\cos \varphi=-1)$. In the system with $m_1=10M_{\rm J}$, $E_{J,\min}=E_{J}(\cos \varphi=-1)$, while $E_{J,\min}=E_{J}(\cos \varphi=1)$ in the systems with $m_1=5M_{\rm J}$ and $m_1=1M_{\rm J}$, so the densities around $E_{J,\min}$ are different in the three systems.

\bibliography{mybib}{}
\bibliographystyle{aasjournalv7}

\end{document}